\documentclass[twocolumn,showpacs,preprintnumbers,amsmath,amssymb]{revtex4}

\usepackage{graphicx,subfigure}
\usepackage{dcolumn}
\usepackage{bm}
\usepackage{hyperref}
\usepackage{textcomp}
\usepackage{nicefrac}
\usepackage{longtable}
\usepackage{mathptmx}
\usepackage{scrtime}
\usepackage{epstopdf}
\usepackage{color}
\usepackage[color=red!40,linecolor=red,textsize=scriptsize]{todonotes}

\newcommand{\mT}{\, \mathrm{mT}}                    
\newcommand{\nm}{\, \mathrm{nm}}                  
\newcommand{\fs}{\, \mathrm{fs}}                   
\newcommand{\ps}{\, \mathrm{ps}}
\newcommand{\ns}{\, \mathrm{ns}}
\newcommand{\GHz}{\, \mathrm{GHz}}

\newcommand{\mum}{\, \mbox{\textmu}\mathrm{m}}

\newcommand{\K}{\, \mathrm{K}}
\newcommand{\degree}{\ensuremath{^\circ}}
\newcommand{\MS}{\ensuremath{M_\mathrm{S}}}                   
\newcommand{\muMS}{\ensuremath{\mu_0 M_\mathrm{S}}}    
                    
\newcommand{\muHext}{\ensuremath{\mu_0 H_\mathrm{ext}}}

\begin{document}

\title{Generation of magnonic spin wave traps}

\author{Frederik Busse}
\affiliation{ I. Physikalisches Institut, University of G\"{o}ttingen, Friedrich-Hund-Platz 1, 37077~G\"{o}ttingen, Germany}
\author{Maria Mansurova}
\affiliation{ I. Physikalisches Institut, University of G\"{o}ttingen, Friedrich-Hund-Platz 1, 37077~G\"{o}ttingen, Germany}
\author{Benjamin Lenk}
\affiliation{ I. Physikalisches Institut, University of G\"{o}ttingen, Friedrich-Hund-Platz 1, 37077~G\"{o}ttingen, Germany}
\author{Marvin Walter}
\affiliation{Institut f\"{u}r Physik, Greisfwald University, Felix-Hausdorff-Straße 6, 17489 Greifswald, Germany }
\affiliation{ I. Physikalisches Institut, University of G\"{o}ttingen, Friedrich-Hund-Platz 1, 37077~G\"{o}ttingen, Germany}
\author{Markus M\"{u}nzenberg}
\email{markus.muenzenberg@uni-greifswald.de}
\affiliation{Institut f\"{u}r Physik, Greisfwald University, Felix-Hausdorff-Straße 6, 17489 Greifswald, Germany }
\affiliation{ I. Physikalisches Institut, University of G\"{o}ttingen, Friedrich-Hund-Platz 1, 37077~G\"{o}ttingen, Germany}

\begin{abstract}
Spatially resolved measurements of the magnetization dynamics induced by an intense laser pump-pulse reveal that the frequencies of resulting spin wave modes depend strongly on the distance to the pump center. This can be attributed to a laser generated temperature profile. On a CoFeB thin film magnonic crystal, Damon-Eshbach modes are expected to propagate away from the point of excitation. The experiments show that this propagation is frustrated by the strong temperature gradient.
\end{abstract}



\maketitle

The manipulation of spin wave frequency and propagation characteristics are of great interest for the design of switching devices such as logic gates in the field of spintronics, and the number of studies in this field grows rapidly \cite{Kruglyak2010JPD_Magnonics, Lenk2011PR_Building}. 
The most promising techniques include (i) current-injected magnetic solitons in thin films with perpendicular anisotropy, which could transmit information directly or alternatively be used to selectively influence another spin wave's propagation~\cite{Mohseni2013Spin}, and (ii) a change in the ferromagnet's temperature and therewith its saturation magnetization. 
The latter can either be brought about by direct contact with e.g. a peltier element, which has been demonstrated by Brillouin-Light-Scattering (BLS) experiments on YIG waveguides \cite{Obry2012_Spin}, or it can be optically induced: The authors of a recent study~\cite{Kolokoltsev2012JAP_Hot} were able to show that by punctually heating up the signal conducting stripline in their network analyzer configuration by up to $\Delta T=70\K$ using a focused cw laser, the magnetostatic surface spin waves propagating along the stripline could be trapped in the resulting potential well. 
In this letter, we address the generation of a spin wave trap on a magnonic crystal by means of a temperature gradient induced by intense laser pulses.

In contrast to the experiments mentioned above, rich magnetization dynamics can be produced without any need for direct contact with the sample by using short optical pulses. 
One approach is using the inverse Faraday effect, which in combination with a spatially shaped pump spot can create propagating droplets of backward volume magnetostatic waves \cite{Satoh2012Directional}. 
On the other hand, the technique applied in this work relies on a thermally induced anisotropy field pulse in the sample to induce magnetization oscillations. 
A common method to access these dynamics, described by the Landau-Lifshitz model of magnetization precession, makes use of the magneto-optical Kerr effect (MOKE) \cite{Lenk2010PRB_Spinwave}. 
Both temporal and spatial information can be obtained by applying time resolved scanning Kerr microscopy (TRSKM). 
Using this technique, propagating spin wave modes have been observed by focusing pump pulses with a full width half maximum (FWHM) of only $10\mum$ on a thin Permalloy film \cite{Dvornik2013PRL_Direct}. 
The spin wave spectrum originating from such optical excitation is usually quite broad: Ultrafast demagnetization leads to a dense population of high energy excitations which then gradually decays into lower energy spin wave modes on a timescale of a few picoseconds. 
Energy transfer from high frequency to low frequency spin waves after excitation by short microwave pumping pulses has been systematically studied by Brillouin-Light-Scattering and it was shown that this mechanism leads to the formation of Bose-Einstein condensates if the pumping is strong enough \cite{Demidov2008Observation}. 
The result is an overpopulation of the lowest energy states which on a continuous film are given by the uniform precession or Kittel mode and by a series of perpendicular standing spin waves. 
Using microstructured magnetic films, so-called magnonic crystals, energy is also transferred into a Damon-Eshbach type mode whose frequency can be tuned in a wide range by choosing appropriate lattice parameters~\cite{Ulrichs2010APL_Magnonic}. 

In this work, we use CoFeB as the sample material due to its low Gilbert damping and high saturation magnetization. 
Ultrashort laser pulses from a regeneratively amplified Ti:Sapphire system are used to (i) excite the magnetization dynamics, (ii) probe the magnetic response of the magnonic crystal, and (iii) create a spin wave trap / resonator.

\begin{figure}
\centering
\subfigure{
\includegraphics[width=0.48\textwidth]{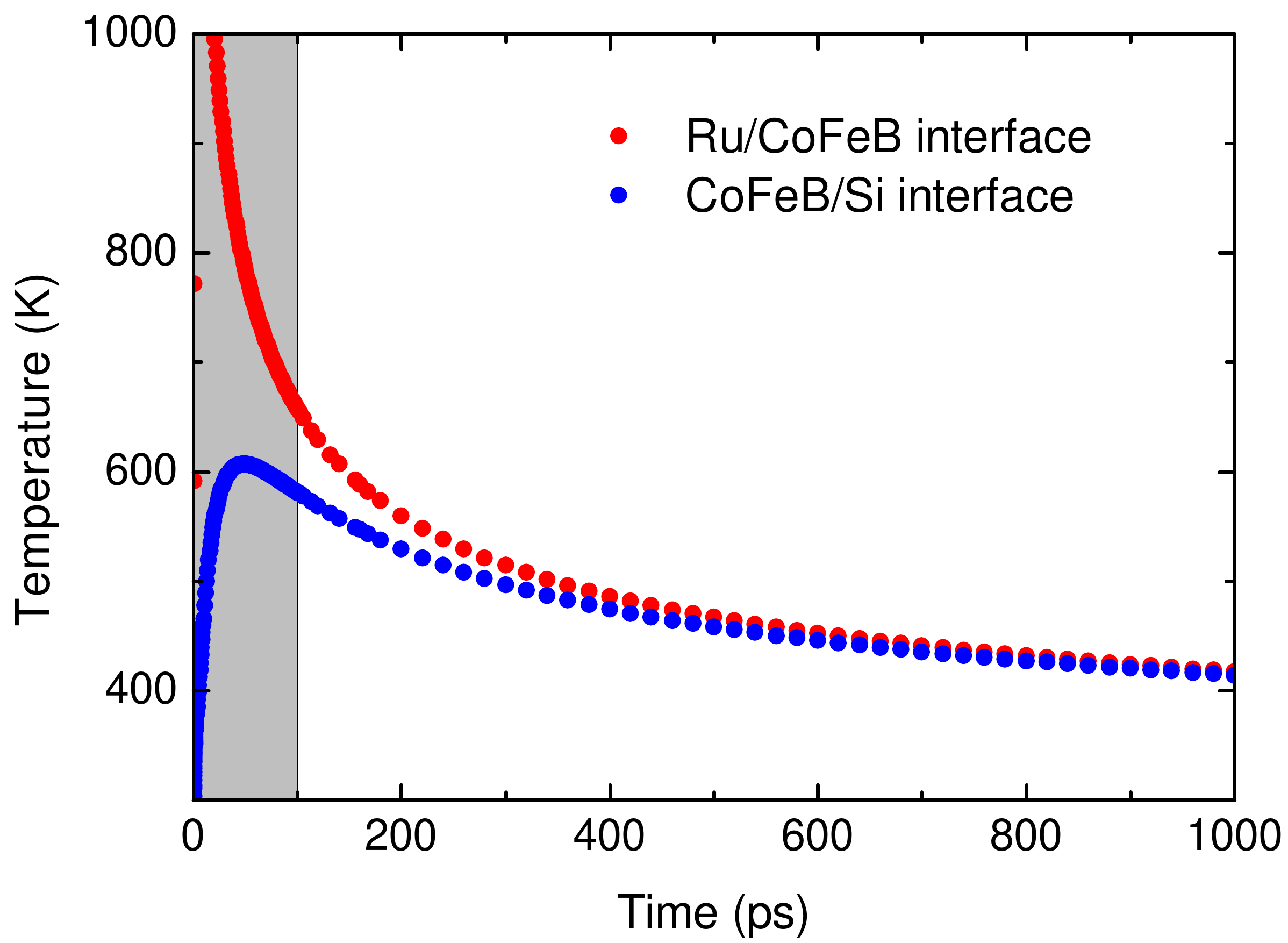}
\label{fig:Simulations:a}
}
\subfigure{
\includegraphics[width=0.48\textwidth]{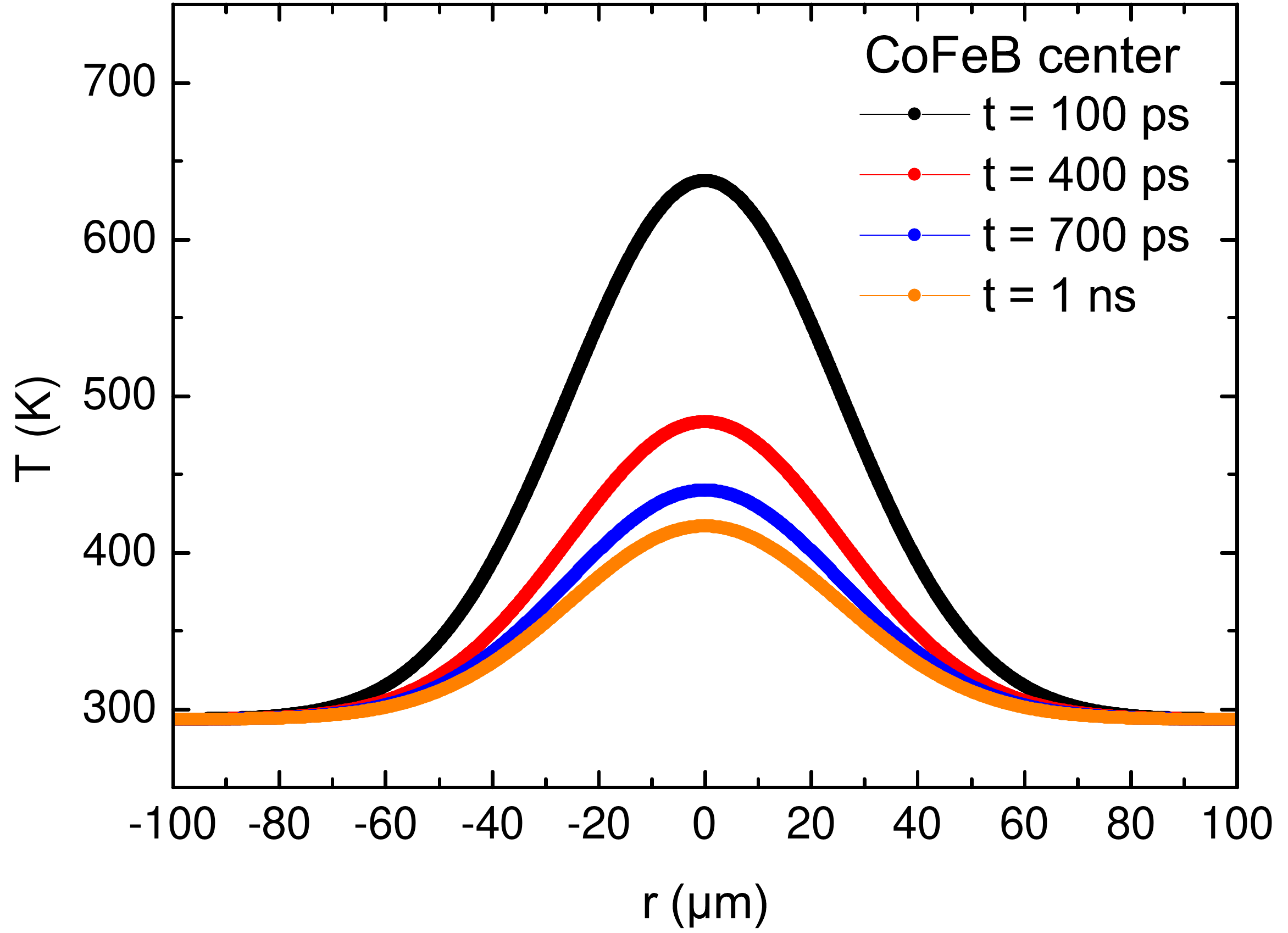}
\label{fig:Simulations:b}
}
\caption{Simulated time evolution of the sample's temperature for different depths~(top) and lateral temperature distribution for different moments in time~(bottom).}
\label{fig:Simulations}
\end{figure}

The software package \emph{COMSOL} has been used to calculate the thermal response of a thin a metallic film to ultrafast laser excitation.
The sample system for these calculations consisted of $3\nm$ of ruthenium capping a $50\nm$ cobalt-iron-boron (Co$_{20}$Fe$_{60}$B$_{20}$) magnetic film on a Si(100) substrate.
The heat diffusion equation,
\begin{equation*}\label{eq:heat-diffusion}
\rho c_p \frac{\partial T}{\partial t} = \nabla (\kappa \nabla T)+Q,
\end{equation*}
is solved in rotational symmetry for isolating sample edges and a fixed temperature at the bottom of the substrate using the material parameters listed in table~\ref{tab:parameters}.
\begin{table}[ht!]
\centering
\begin{tabular}{lcccc}
\hline
Material	& $\rho$ (kg$\,$m$^{-3}$) & $c_p$ (J$\,$kg$^{-1}\,$K$^{-1}$) & $\kappa$ (W$\,$m$^{-1}\,$K$^{-1}$) & $R$ \\ \hline \hline
Ru 							& 12370 \cite{Walter2011NM_Seebeck}	 & 238 \cite{Walter2011NM_Seebeck}		& 117 \cite{Walter2011NM_Seebeck} 	 & 0.70 \cite{Hass1981}		        \\
Co$_{20}$Fe$_{60}$B$_{20}$ 	 & 7700 \cite{OHandley1976}	& 440 \cite{Walter2011NM_Seebeck}		& 87 \cite{Walter2011NM_Seebeck} 	& 0.72 \cite{Johnson1974}		\\
Si 							& 2330 \cite{Enghag2004}	& 712 \cite{Enghag2004}		& 153 \cite{Enghag2004}	& -   \\ \hline
\end{tabular}

\caption{Material parameters of the \emph{COMSOL} simulation for 3\,$nm$~Ru~/~50\,$nm$~CoFeB~/~50\,$\mu$m~Si sample: Density $\rho$, heat capacity $c_p$, and thermal conductivity $\kappa$ and reflectivity $R$ at $\lambda=800\,$nm. CoFeB values $\rho$ and $c_p$ are average values for Co and Fe, CoFeB reflectivity is approximated by the value for Co.}
\label{tab:parameters}
\end{table}
Starting from equilibrium at room temperature, energy is deposited by an ultrashort laser pulse with a duration of $50\fs$.
The optical penetration depth is $\Lambda=16.1\nm$ in accordance with the value for ruthenium~\cite{Palik1985} as well as with the average value of cobalt and iron, respectively~\cite{Johnson1974}.
In the film plane, a Gaussian intensity profile is assumed with a FWHM of $60\mum$. The energy carried by each pulse amounts to a total of $1.6 \mu J$, as will be the case in the experiments presented below.
The results of the simulation are shown in Fig.~\ref{fig:Simulations}: In the beginning, the laser pulse produces a sudden rise in temperature.
After thermalization of the optically excited electrons and equilibration of the spin and phonon subsystems, known to take place on timescales of $\approx 100\,$fs and $\approx 1\,$ps respectively, the modeling yields an effective sample temperature, i.e the temperature of the magnetic system. 
During the first $\approx 100\,$ps the spatial as well as the temporal heat gradient are rather large, whereas at later times the temperature remains at a high mean value and only a negligible depth profile remains for most part of the CoFeB film.

While the temperature is mainly homogeneous throughout the sample depth, it changes significantly across its plane, as shown in Fig.~\ref{fig:Simulations} (bottom). 
The Gaussian distribution of laser intensity in the pump spot produces a temperature profile that persists longer than the lifetime of the observed coherent spin wave modes. 
During this time (up to 1\,ns), no significant heat transport takes place on a micrometer scale and the FWHM of the lateral temperature distribution remains unchanged. 
In accordance with the Curie-Weiss law, the temperature increase quenches  sample's saturation magnetization so that a potential well is formed which effectively prevents the escape of spin waves from this region.

Magnetization dynamics experiments were conducted on amorphous $50\nm$-thick Co$_{40}$Fe$_{40}$B$_{20}$ films magnetron-sputtered onto a Si(100) substrate and capped with a $3\nm$ Ru layer to prevent oxidation. 
Ultrashort laser pulses (central wavelength $\lambda_c=800\nm$, pulse duration $50\fs$)  amplified by a Coherent RegA 8040 regenerative amplifier were used to excite and detect the magnetization dynamics in a pump-probe experiment as described in ref.~\cite{Djordjevic2006JAP_Intrinsic}.
The experimental parameters are analogous to those in the presented COMSOL simulation. 
Additionally, an external magnetic field $\muHext=\pm130\mT$ is applied at $20^\circ$ with respect to sample plane. 
The experiments were performed  separating the pump and probe spots on the sample and measuring the magnetization dynamics as a function of pump-probe distance, thus allowing us to determine the shift in magnetization oscillation frequency along the temperature gradient (i.e the spin wave well).
\begin{figure*}
\centering
\subfigure{
\includegraphics[width=0.15\textwidth]{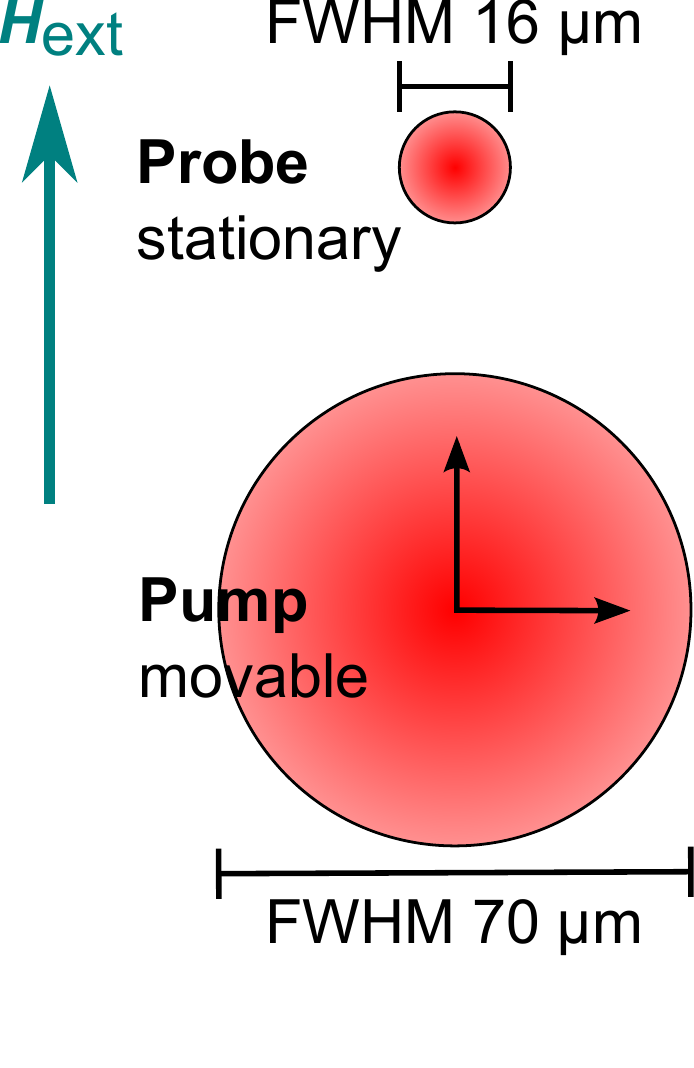}
} \quad \subfigure{
\includegraphics[width=0.37\textwidth]{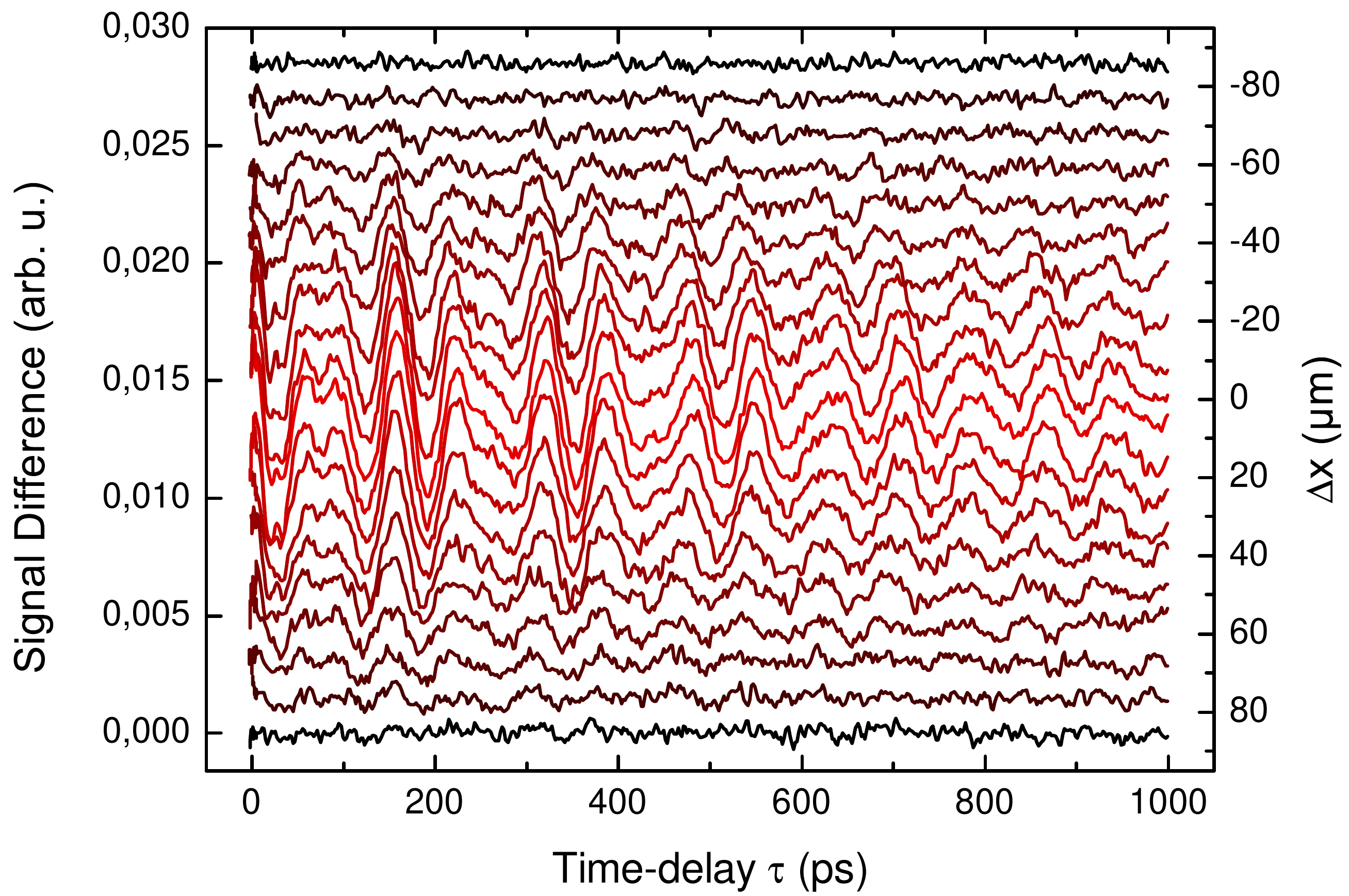}
} \subfigure{
\includegraphics[width=0.37\textwidth]{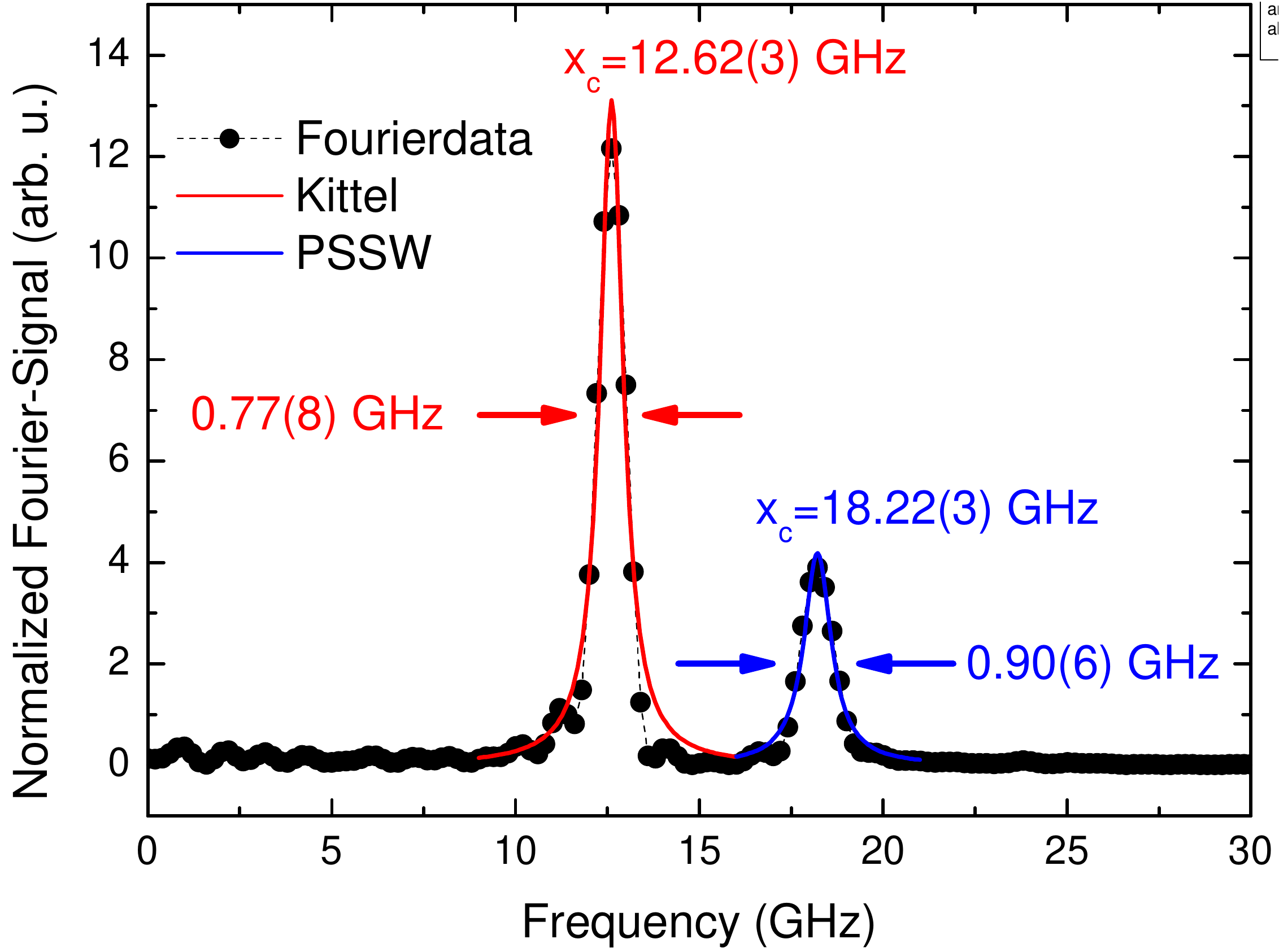}
}
\caption{Analysis of TRMOKE data: On the left side, the pump-probe geometry and the direction of the external magnetic field are shown. After calculating the difference of the corresponding time-resolved traces $\Delta M(\tau)$ acquired at $\muHext=\pm130\mT$ and subtraction of the incoherent background, a coherent oscillation of the magnetization is visible~(center). In the frequency domain~(right), two modes occur and are identified as the uniform precession (Kittel $k=0$ mode) and as perpendicular standing spin waves (PSSW), respectively.}
\label{fig:DataAnalysis}
\end{figure*}

Using variable time delay $\Delta\tau$ between pump and probe pulses, the time-resolved magneto-optical Kerr effect (TRMOKE) reveals  magnetization precession on timescales of up to $1\ns$, which changes phase by $\pi$ between positive and negative (i.e. reversed) field directions.
For the quantitative analysis, the difference between both field directions is calculated. 
An incoherent background remains which originates from high frequency and high-$k$ magnons excited by the intense pump beam ~\cite{Lenk2010PRB_Spinwave}.
After respective subtraction, a fast Fourier transform is performed and the resulting peaks in frequency domain can be analyzed (see figure \ref{fig:DataAnalysis}).

The dataset presented in Fig.~\ref{fig:DataAnalysis} has been obtained on a continuous CoFeB reference film of thickness~$d=50\nm$. Two modes of magnetic precession are observed. Based on earlier results, these can be identified as the in-phase precession of all spins (uniform Kittel mode) at $12.6 \GHz$ and a first order (i.e.\ $n=1$) standing spin wave with wave vector~$k=n \pi \, d^{-1}$ perpendicular to the sample plane~(PSSW) at $18.2 \GHz$~\cite{Ulrichs2010APL_Magnonic,Lenk2010PRB_Spinwave}.
Both Kittel and PSSW modes have no wave vector components in the lateral direction. In other words, they do not propagate on the sample but have a rather localized character at the spot of (optical) excitation.
Consequently, spatially resolved measurements should show no significant precession outside of the pump laser spot.

\begin{figure}[h!]
\caption{Experimental results on a continuous film. In~(top), open squares represent fitted peak positions. The precession frequency observed after optical excitation is not constant across the pump spot. For the Kittel mode, a local magnetization can be calculated. Together with the magnetization curve shown in the inset, the temperature of the spin system can be derived~(bottom). Closed diamonds correspond to a displacement of pump and probe spots orthogonal to the applied field, open squares depict a parallel displacement. Solid lines are guides to the eye. Curves are offset so that the frequency dip of the Kittel mode is centered at $\Delta x=0$.}%
\vspace{0.6em}
\centering
\subfigure{
\includegraphics[width=0.31\textwidth]{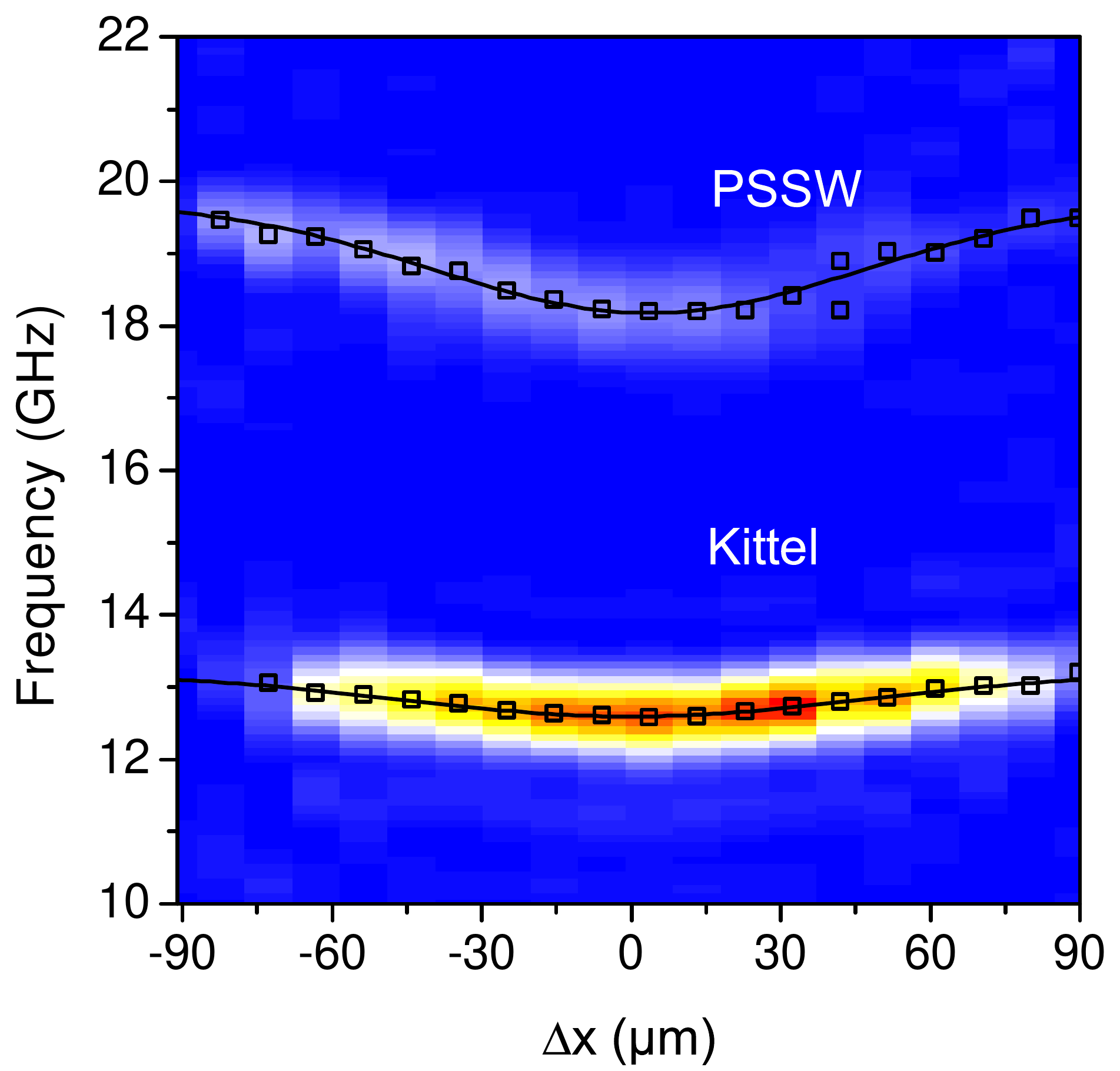}
\label{fig:Conti-Position-Dependence:a}
} 
 \subfigure{
\includegraphics[width=0.35\textwidth]{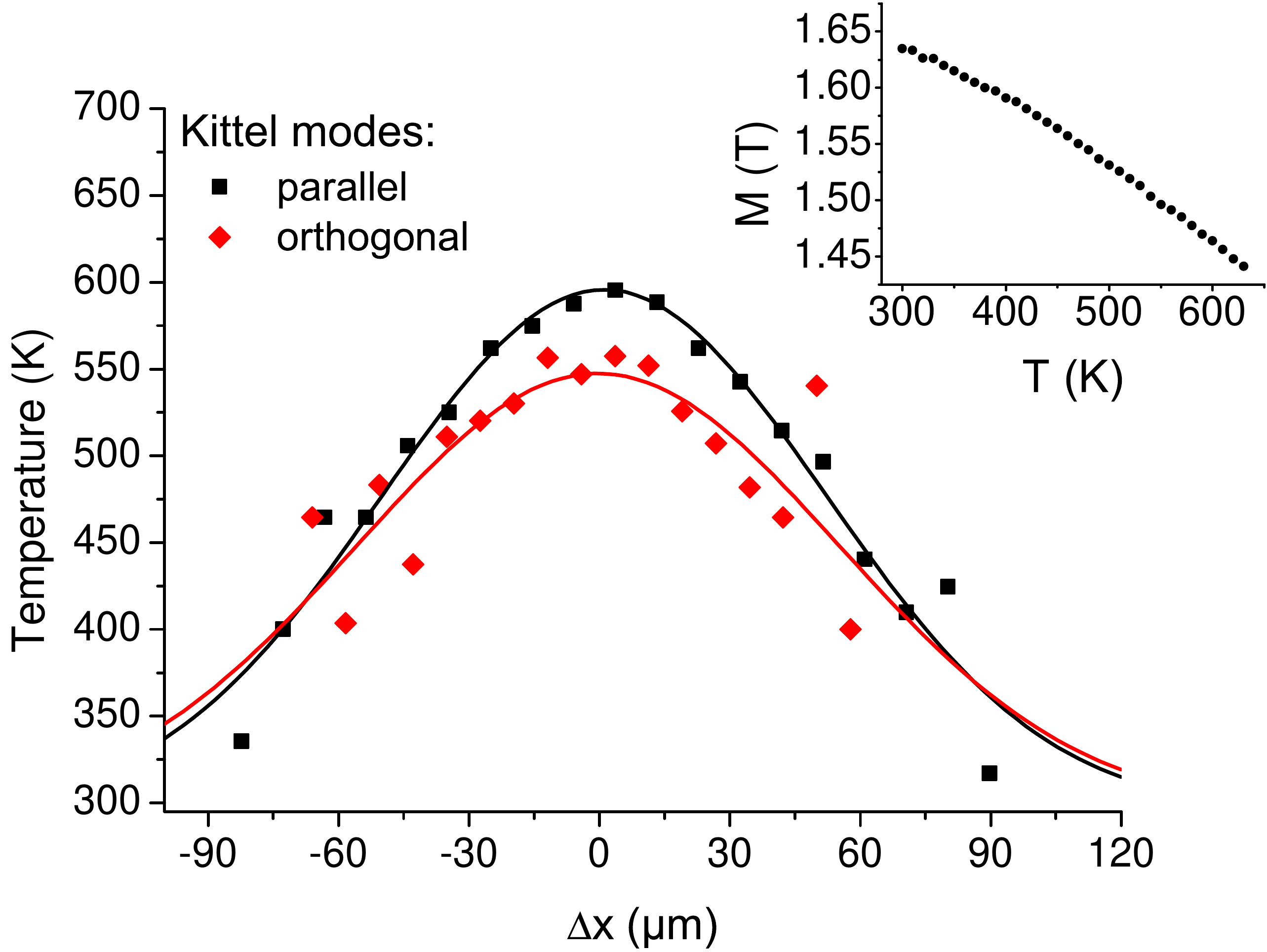}
\label{fig:Conti-Position-Dependence:b}
}
\label{fig:Conti-Position-Dependence}
\end{figure}

Fig.~\ref{fig:Conti-Position-Dependence} (top)  shows the color-coded Fourier power~$P_\mathrm{FFT}$ of magnetization oscillation as a function of spatial separation~$\Delta x$ between the centers of pump and probe spot. 
In this dataset, pump displacement was performed parallel to the external field direction. 
For better comparison of different measurements, the Fourier power is normalized to the sum of all transformed data points, allowing to see how much the signal stands out against background noise.
On the one hand, the precessional amplitude (represented by the color code) depends on the distance to the center of the pump pulse. 
This is due to the laser intensity profile and the localized character of the observed modes. 
On the other hand, also the frequency is strongly position dependent.
The latter effect can be explained as a consequence of the increased disorder caused by the intense heating, which leads to a decrease in saturation magnetization and therefore to a change of the spin wave spectrum.

Using the theoretical dispersion of Kittel mode in the Landau-Lifshitz formalism,
\begin{equation}\label{eq:kittel-dispersion}
\omega^\mathrm{theo}_K =
\gamma \mu_0 \sqrt{H_x \big(H_x + M_\mathrm{S} \big)},
\end{equation}

the profile in observed frequency can be used to calculate the laser induced temperature increase: In Eq.~(\ref{eq:kittel-dispersion}) the saturation magnetization~$\muMS$ is regarded as the only free parameter, such that $\omega^\mathrm{theo}_K = \omega^\mathrm{theo}_K ( \MS )$ . Comparing this with the experimentally observed frequency profile $\omega^\mathrm{exp}_K ( \Delta x)$ (open squares in Fig.~\ref{fig:Conti-Position-Dependence}(top)), a corresponding profile in magnetization $\MS ( \Delta x)$ is calculated. 
The magnetization profile is then compared to the magnetization curve $M(T)$ obtained for a CoFeB sample of equal thickness and composition using a Vibrating Sample Magnetometer (VSM) (inset in Fig.~\ref{fig:Conti-Position-Dependence}). The resulting position dependent temperature profile is shown in figure~\ref{fig:Conti-Position-Dependence} (bottom).

\begin{figure*}
\centering
\includegraphics[width=0.98\textwidth]{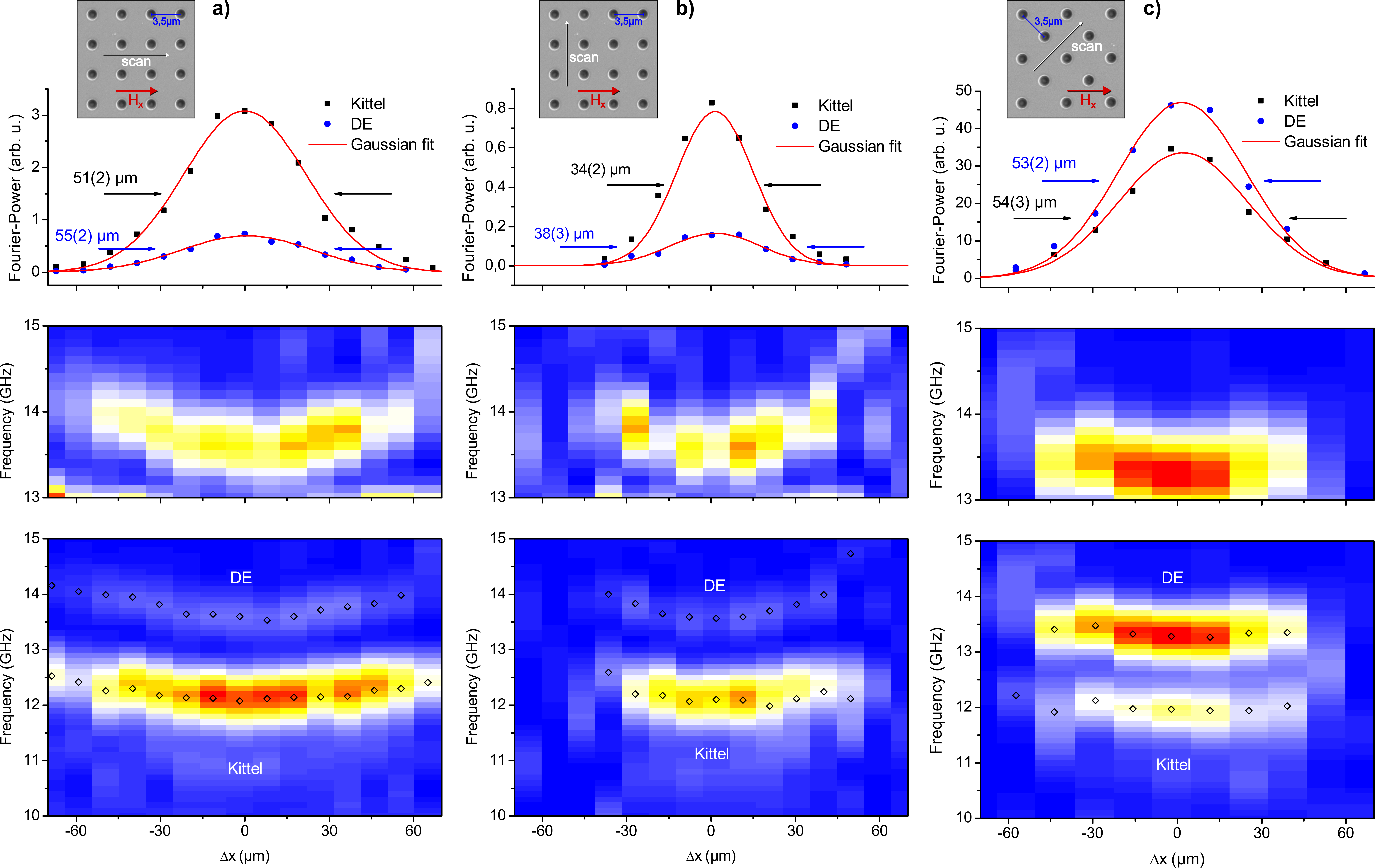}
\caption{Experimental results on a magnonic crystal. Bottom: Fourier-spectrum in analogy to Fig.~\ref{fig:Conti-Position-Dependence}(a) with the additional magnonic Damon-Eshbach (DE) Bloch mode. Above, the DE-mode is shown with an adapted color code for better visibility. The Fourier power for fundamental Kittel mode (squares) and dipolar DE mode (circles) is plotted as a function of the relative position between pump and probe. No significant spatial widening is observed for the DE mode frequency profile. Top: Scanning direction relative to the applied field and to the magnonic crystal.}
\label{fig:PeakHeights}
\end{figure*}

While we expect that the Kittel and PSSW do not propagate across the sample plane, in magnonic crystals composed of periodically arranged antidots the excitation of dipolar Damon-Eshbach surface waves~(DE) of selective wave vector has been shown~\cite{Ulrichs2010APL_Magnonic}. This provides a possibility for information transport via local excitation of propagating spin waves followed by a spatially separated detection.
Magnetization dynamics measurement on a magnonic crystal and its analysis are presented in Fig.~\ref{fig:PeakHeights}. 
In these measurements, pump and probe beam were separated (a) parallel, (b) orthogonal and (c) at an angle of 45\degree{} to the direction of the external magnetic field. 
In contrast to the measurements on the continuous film, an additional (magnonic) Damon-Eshbach mode is visible (bottom images of Fig.~\ref{fig:PeakHeights}). 
As has already been observed above for the Kittel and PSSW modes, the precession frequency of DE mode shows a Gaussian dependence on position with a minimal frequency at the position of maximal pump intensity, which is caused by the increased temperature.
The measurements should reveal a second effect, though: Damon-Eshbach surface waves are known to propagate mainly orthogonal to the direction of the in-plane magnetic field. 
In magnonic antidot crystals, only a DE mode parallel to the direction of the smallest hole-to-hole distance is detected ~\cite{Ulrichs2010APL_Magnonic}.
Since the propagation direction of a dipolar surface wave is rotated by 180\degree{} when the external field is reversed, and because the data shows the difference between the signals measured for positive and negative field direction, a spatial widening of the DE-mode is expected.
From the damping time in the TRMOKE data of $300-800\ps$, the propagation length of spin waves in CoFeB can be estimated to be at least $100\mum$ so that detection of the DE-mode should be possible well outside of the pump spot.

The magnetization oscillation's Fourier power for each measurement is plotted in the top row of Fig.~\ref{fig:PeakHeights}. 
Solid lines represent Gaussian fits to the experimental points, the fitted widths amount to around $50\mum$. 
By comparison of the (localized) Kittel and the (propagating) DE mode, the surface mode's propagation characteristics can be analyzed. 
Since both modes show the same FWHM, it can be concluded that no propagation occurs into the direction of the external field (Fig. \ref{fig:PeakHeights}(a)), as could be presumed. 
Peculiarly, for the orthogonal and 45\degree{} configuration there is no significant broadening of the DE mode either, meaning that there is no propagation out of the excitation spot.

Two damping mechanisms can be considered to explain the observed behavior: On the one hand, the pump pulse creates a magnon population of very high density, rendering the picture of ballistic spin wave propagation invalid. 
Instead, intense scattering takes place that results in a strong overall damping.
On the other hand, a spin wave travelling away from the spot of excitation would propagate towards an increasing effective saturation magnetization due to the heat gradient imposed by the pump laser. 
As we have shown, this change in saturation magnetization drastically impacts the supported frequency, and consequently must result in repeated scattering of the DE-magnons. 

The presented experiments and their analysis carry two important points: Firstly, an effective spin-wave well is formed by the local absorption of the optical pump pulse. 
In analogy to \cite{Kolokoltsev2012JAP_Hot} we observe a magnetization profile $\MS (x)$ that follows the intensity profile of optical excitation and strongly influences the observed spin wave spectrum.
Despite the ultrashort character of the excitation, the temperature profile remains in effect over the complete range of observed time delays, namely up to $1\ns$.
In view of magnonics and their applications, a possible scenario is an optical lattice on a continuous magnetic film. Effectively, a dynamic magnonic crystal can be created in this way without limitations by lithography.

Secondly, we observed the absence of spin wave propagation away from the excitation spot, which is mainly caused by two distinct mechanisms: As discussed in references \cite{Djordjevic2007PRB_Connecting, Lenk2010PRB_Spinwave, Lenk2011PR_Building}, optical spin wave excitation is highly non-equilibrium. 
The resulting spin wave density is far above the ballistic limit, thus leading to a high probability for scattering between spin waves and a drastically reduced mean free path. 
A temperature gradient imposes additional scattering as spin waves are continuously reflected when entering a colder region with higher saturation magnetization.
This effect might be used to trap spin waves or selectively block their propagation and must certainly be considered when optically exciting propagating surface waves.

\begin{acknowledgements}Maria Mansurova thanks Soham Manni for assistance and discussion during VSM measurements. We thank the German Research Foundation (DFG) for funding through MU 1780/ 6-1 Photo-Magnonics, SPP 1538 SpinCaT and SFB 1073.
\end{acknowledgements}

\end{document}